\documentclass[twocolumn]{article}
\usepackage[left=2.5cm, right=2.5cm,top=2cm,bottom=2cm]{geometry}

\usepackage{xcolor}
\usepackage{amsmath}
\usepackage{amsfonts}
\usepackage{amssymb}
\usepackage{graphicx}
\usepackage{hyperref}
\usepackage{multirow}
\usepackage{authblk}
\usepackage{tikz}
\usepackage{dblfloatfix}
\usepackage[per-mode = symbol]{siunitx}
\usepackage[normalem]{ulem}
\usepackage[font=footnotesize,labelfont=bf]{caption}
\usepackage{endnotes}
\usepackage{tabularx,booktabs}

\usepackage[style=phys,backend=bibtex]{biblatex}
\title{Fragility of Aeolian Ripples}
\author[1]{Constantin Rein}
\author[1]{Katharina Tholen}
\author[2]{Lior Saban}
\author[2]{Itzhak Katra}
\author[3]{Hezi Yizhaq}
\author[1]{Klaus Kroy}
\affil[1]{\small Leipzig University, Institute for theoretical Physics, Faculty of Physics and Earth System Sciences, Germany}
\affil[2]{\small Department of Environmental, Geoinformatics and Urban Planning Sciences, Ben-Gurion University of the Negev, Be’er Sheva, Israel}
\affil[3]{\small Department of Solar Energy and Environmental Physics, Ben-Gurion University of the Negev, Be’er Sheva, Israel}

\begin{document}
\twocolumn[
    \begin{@twocolumnfalse}
\maketitle
  \begin{abstract}
Sand surfaces exposed to the wind are covered by aeolian sand waves of diverse wavelengths, from centimetres for small ripples to decameters for dunes. Sand dunes can be said to be antifragile, since their growth is not prohibited  by storms, whereas ripples and megaripples are fragile since they get flattened. The fluid-entrainment hypothesis attributes their fragility to a common cause: the spontaneous aerodynamic entrainment of grains. Here, we challenge this conventional view by revealing its major shortcomings and inconsistencies. An alternative framework, routed in the granular bed rheology, posits that impact ripples are erased when the two-phase flow of ballistically hopping grains on a solid sand surface gives way to a three-phase flow, featuring an intermediate, collision-dominated, fluid-like transport layer. Notably, both martian wind-drag ripples and terrestrial aerodynamic ripples, which emerge via a hydrodynamic instability, are insensitive to such ``surface-melting''. Their destruction is predicated on the transition from aerodynamically smooth to rough conditions, instead. Our theoretical arguments are well supported by a stability diagram in terms of Shields number and grain diameter, assembled from a carefully reassessed large body of literature, and backed up by dedicated control measurements.
  \end{abstract}
  \vspace{\baselineskip}
\end{@twocolumnfalse}
]

\begin{figure}[ht]
    \centering
\includegraphics[width=\linewidth]{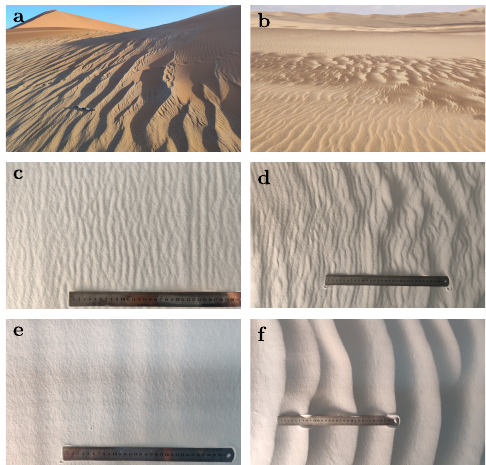}
\caption{\textbf{Aeolian sand patterns in the field and in wind tunnels.} \textbf{(a-b)}: Co-existence of dunes and polymorphous ripple patterns in the field, illustrated for the Sossusvlei \textbf{(a)} and Torra Bay \textbf{(b)} areas, Namibia. \textbf{(c-f)}: Co-evolution of ripples in the wind tunnel (scale bar $33\unit{\centi\metre}$). For $u_* = 0.35 \unit{\metre\per\second}$,  impact ripples develop \textbf{(c)} after around 4 minutes of wind exposure. They morph into an aerodynamic ripple pattern with superimposed impact ripples   \textbf{(d)} half an hour later. For $u_* = 0.48 \unit{\metre\per\second}$, impact ripples are absent both  \textbf{(e)} before (at around one minute of wind exposure) and \textbf{(f)} after (around 22 minutes later) aerodynamic ripple formation.}
\label{fig:ripple-evolution-images}
\end{figure}

Atmospheric conditions shape unique wind-blown sand patterns on Earth and on extraterrestrial bodies. Most abundant are ripple patterns as exemplified in Tab.~\ref{tab:ripples} and Fig.~\ref{fig:ripple-evolution-images}(a,b). They decorate nearly any wind- or water-exposed sand-surface, including that of their giant relatives, sand dunes. After decades of ripple research, the challenge to resolve the grain-scale mechanisms underlying ripple formation persists \cite{lester_emergence_2025,Tholen24,Lapotre2016}. This study shifts the focus from ripple formation  to the poorly elucidated phenomenon of ripple disappearance (Fig.~\ref{fig:ripple-evolution-images}). We thereby redirect the attention from the many similarities between ripples and dunes to what is arguably the most profound distinction between them: ripples can be destroyed by strong winds but dunes not. They simply migrate and, depending on sand supply, grow or shrink faster, under stronger winds \cite{kroy_minimal_2002} so that they can be said to be ``antifragile'' \cite{taleb_antifragile}. In contrast, ripples are observed to disappear under strong winds \cite{Bagnold41,Walker81,Schmerler2016,chepil1945dynamics,goossens_aeolian_1991,LingYuquan1998,Sharp63,LingYuquan2003,isenberg_2011,Katra2014,yizhaq_evolution_2012,yizhaq_longevity_2015}, and should be categorized as ``fragile''.

\begin{table*}[ht]
\centering
{\footnotesize
\begin{tabularx}{\textwidth}{@{} l X X X X @{}}
\toprule
 & Impact ripples & Aerodynamic ripples & Mars wind-drag ripples& Megaripples \\
\midrule
Traits  & $\lambda \simeq \unit{\centi\metre}$, unimodal sand
                & $\lambda \simeq \unit{\deci\metre}$, unimodal sand
                & $\lambda \simeq \unit{\metre}$, unimodal sand & $\lambda \simeq \unit{\deci\metre}$, bimodal sand \\ \addlinespace 
Formation & Saltation impacts on static bed
          & Aerodyn. instability \cite{Tholen24}
          & Aerodyn. instability \cite{Lapotre2016}
          & Saltating fine, reptating coarse grains \cite{tholen_megaripple_2022} \\ \addlinespace
Flattening & Surface melting
           & Aerodyn. roughness
           & Aerodyn. roughness 
           & Coarse-grain saltation \\ \addlinespace 
Photos & \includegraphics[width=\linewidth]{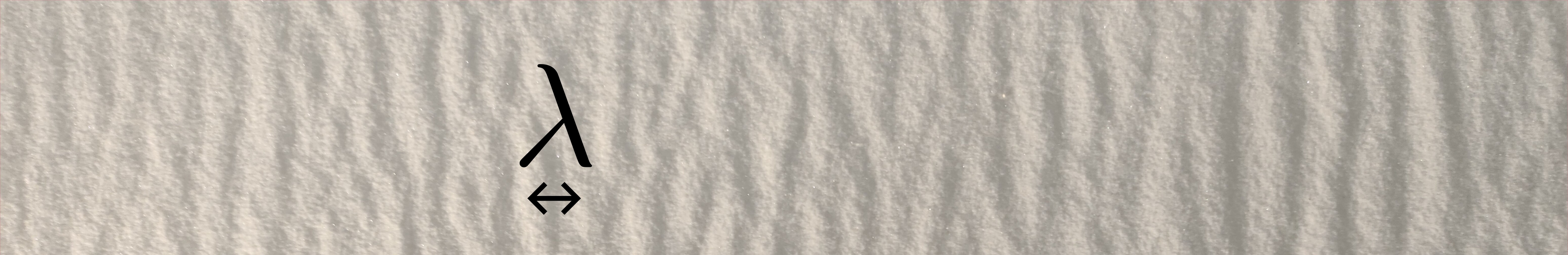} & \includegraphics[width=\linewidth]{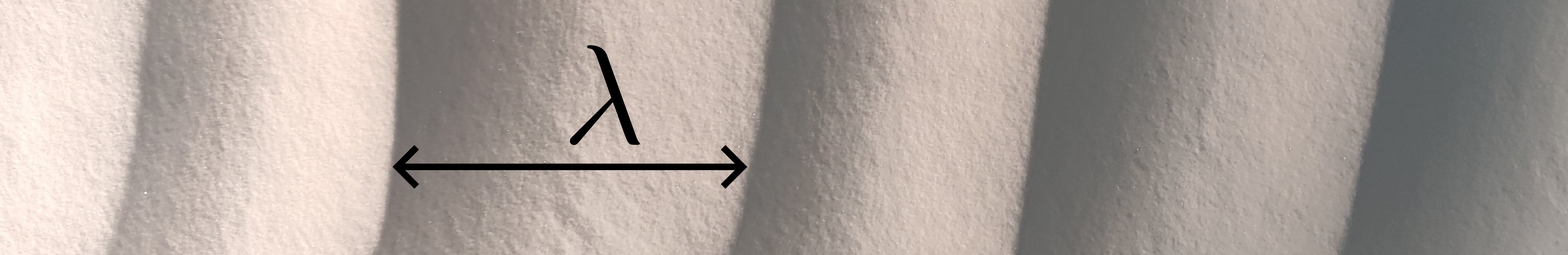} & \includegraphics[width=\linewidth]{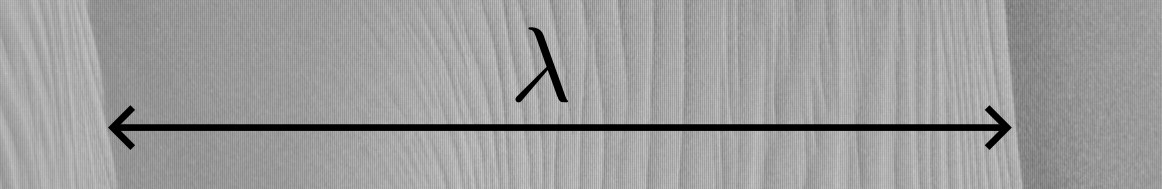} &\includegraphics[width=\linewidth]{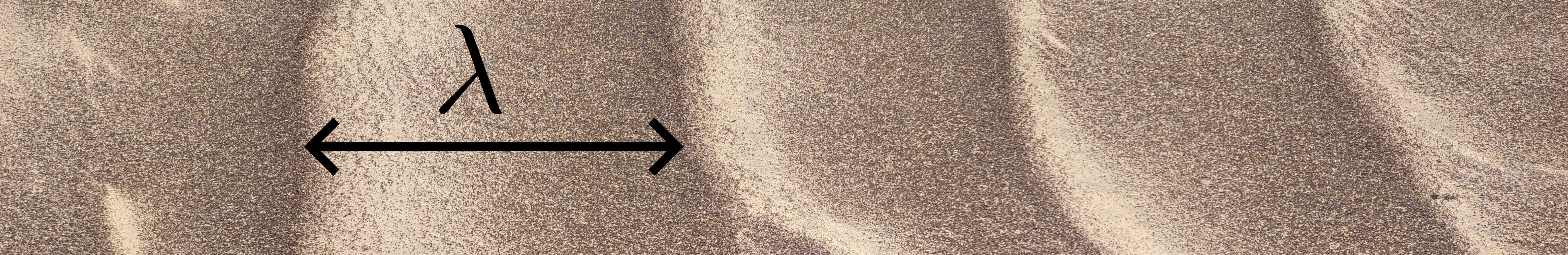} \\
\bottomrule
\end{tabularx}
}
\caption{\textbf{Aeolian ripple formation and flattening.} Wind-drag ripple photo from Mars Rover \cite{nasa-wind-drag-ripples}, for the others see Fig.~\ref{fig:ripple-evolution-images}}
\label{tab:ripples}
\end{table*}

The fragility of aeolian ripples, especially of so-called impact ripples, is not yet well understood. Previous explanations of impact ripple erasure were generally based on a hydrodynamic feedback mechanism: the rippled topography induces an increased wind shear stress at the ripple crests, supposed to limit further growth \cite{sullivan_broad_2020,Bagnold41,manukyan_formation_2009,Schmerler2016,Bar16,Siminovich19} and to eventually cause its own flattening. More precisely, this was suggested to happen once the fluid-entrainment threshold (the wind shear stress needed to initiate sand transport by fluid drag \cite{ShaoLu_2000,Bagnold41}) is surpassed \cite{Siminovich19,Bar16}.

Since this conventional fluid-entrainment argument is applicable to any local bed elevation, so-called megaripples, characterized by their unique, co-evolving bimodal grain size distribution\cite{tholen_megaripple_2022,manukyan_formation_2009,Laemmel2018}, were hypothesized to disappear by the same mechanism \cite{Katra2014,yizhaq_evolution_2012,isenberg_2011,yizhaq_longevity_2015,manukyan_formation_2009,Bagnold41,sullivan_broad_2020,yizhaq_morphology_2008}, although a mechanistically refined, alternative explanation was put forward \cite{Katra2014,yizhaq_longevity_2015,john_m_ellwood_peter_d_evans_an_small_1975,yizhaq_morphology_2008,yizhaq_evolution_2012,isenberg_2011}. 
Unfortunately, as outlined below, hypothetical ripple destruction mechanisms based on the fluid-entrainment argument have serious inconsistencies and turn out to be incapable of predicting concise ambient conditions for the vanishing of impact ripples, let alone megaripples.  

Arguably, the vagueness of existing explanations can be blamed on the lack of a comprehensive theory for ripple formation and the lack of systematic literature data on their disappearance. New valuable insights can therefore be gained from the recent systematic demonstration of the co-existence of impact ripples and larger-scale aerodynamic ripples \cite{Miller1987,Tholen24, Lapotre2016,alvarez_ripples_2025} and its mechanistic explanation in terms of a competition of two distinct formation mechanisms, first for Mars \cite{Lapotre2016}, and then for Earth \cite{Tholen24}. 
Due to this very recent update of the conventional ripple classification scheme, existing literature data has to be carefully reviewed. We now know that it is crucial to distinguish impact and aerodynamic ripples before drawing conclusions concerning their disappearance. In fact, the failure to do so may account for the large scatter in their previously reported  characteristics, which possibly hints at their distinct growth mechanisms.

In the following, we contrast the conventional fluid-entrainment criterion for ripple flattening with a competing ``surface-melting hypothesis'', based on recent advances in the understanding of aeolian sand transport \cite{Paehtz_unification_2020,Paehtz_unified_2021,Pahtz_scaling_laws_2023,Tholen_scaling_2023,carneiro_midair_2013,ralaiarisoa_transition_2020}. Similarly as the mentioned mechanistic criterion for megaripple flattening, this new hypothesis sharpens our understanding of the grain-scale mechanisms responsible for ripple fragility. We condense the emerging consistent theoretical picture of aeolian mesoscale bedforms into a ripple stability diagram for impact ripples, aerodynamic ripples, and megaripples and verify it by pertinent literature data and our own dedicated measurements.

\section*{Results}

\subsubsection*{Deconstruction of the conventional view}

\begin{figure*}[ht]
    \centering
    \includegraphics[width=\linewidth]{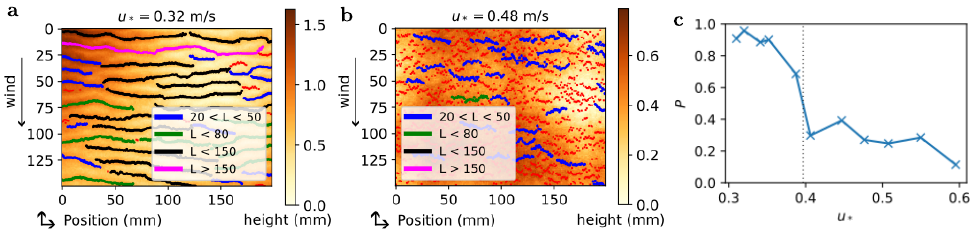}
\caption{\textbf{Disintegration of impact ripples under strong winds in wind tunnel measurements:} crest line plots of the impact-ripple wave pattern for two wind shear velocities $u_* = 0.32 \unit{\metre\per\second}$ \textbf{(a)} and $u_* = 0.48 \unit{\metre\per\second}$ \textbf{(b)}; colorbars indicate height in $\unit{\milli\metre}$, the length $L$ of crest lines is color-coded. A drastic drop in the order parameter $P$, the fraction of connected crest lines, marks the vanishing of the impact ripple pattern \textbf{(c)} at $u_* \approx 0.4 \unit{\metre\per\second}$ (dotted line). Its abrupt disintegration evokes the notion of a dynamical phase transition rather than a gradual suppression of coherent wave trains.}
\label{fig:crestline_plots_w_clusters}
\end{figure*}

The conventional fluid-entrainment hypothesis is built on the notion that the wind erodes incipient ripples when the surface shear stress exceeds the fluid-entrainment threshold. The latter presumably marks a change in the mobilization mechanism of grains: beyond the threshold shear velocity $u_{*\text{f}}$, they can be lifted from the static bed, merely by fluid drag forces \cite{Bagnold41,ShaoLu_2000}, without help from impacting other grains. Topographically induced shear stress peaks at the crests should accordingly lead to crest erosion and growth saturation \cite{Siminovich19,Bar16,sullivan_broad_2020,Bagnold41,manukyan_formation_2009,Schmerler2016}.

One of the shortcomings of this intuitive explanation is revealed by its applicability to an arbitrary surface undulation. Consequently, the fluid-entrainment mechanism was hypothesized to be responsible for the deconstruction of both impact ripples and megaripples \cite{isenberg_2011,Katra2014,yizhaq_evolution_2012,yizhaq_longevity_2015,manukyan_formation_2009,Bagnold41,yizhaq_morphology_2008,sullivan_broad_2020}. But field observations indicate that, upon megaripple flattening, smaller-scale ripple structures may survive, which are believed to be impact ripples \cite{isenberg_2011,yizhaq_evolution_2012,yizhaq_longevity_2015,yizhaq_morphology_2008,Bagnold41,john_m_ellwood_peter_d_evans_an_small_1975}. Rough estimates of the shear velocity $u_*$ reveal that megaripple flattening in the field \cite{isenberg_2011,yizhaq_evolution_2012}, and the associated impact ripple ``survival'', occurs for winds strengths well below the fluid-entrainment threshold $u_{*\text{f}}$ for the coarse megaripple grains. While we are aware of the complexity and diversity of multi-scale ripple structures in polydisperse sands, we believe that the fluid-entrainment hypothesis is in conflict with such observations. It also fails to answer the basic conceptual question: why are megaripples eroded but impact ripples emerge and survive? If crest erosion arises due to a hydrodynamic feedback to the rippled topography, why should impact ripples not be eroded and suppressed if megaripples are?

A competing argument is based on the characteristic bimodal grain size distribution of megaripples, which reflects a unique bimodal transport mechanism: the coarser grains move only by reptation (short hops) induced by the impacts of the smaller, saltating grains \cite{tholen_megaripple_2022}. Now, as much as megaripple formation relies on this widely accepted delicate bimodal transport mode, megaripple decay can clearly and consistently be associated with its derangement. Namely, large shear stresses, yet well below the coarse (and possibly even fine) grains' fluid-entrainment threshold, eventually allow for saltation of the coarser, previously only reptating, grains. Megaripple destruction in strong winds is thereby attributed to a fundamental change in the transport mode, independent of the fluid-entrainment mechanism.

Attempts to evoke the fluid-entrainment hypothesis for impact ripple vanishing meet further inconsistencies. To begin with, we observed that, in our experiments, impact ripples developed for shear velocities $u_*$ well above the fluid entrainment threshold (see Fig.~\ref{fig:ripple-evolution-images}(c,d)), in line with many literature reports \cite{Bagnold41,Walker81,Schmerler2016,Tholen24,zheng_mechanics_2009,cheng_experimental_2018,Li2003,alvarez_ripples_2025}. 
In fact, since wind fluctuations in wind-tunnels are much constrained, transport there cannot even be initiated by wind drag forces below the fluid-entrainment threshold without artificial sand feeding. Which effectively restricts, by design, the observational wind regime to $u_* \gtrsim u_{*\text{f}}$ in such experiments (\cite{Tholen24,Cornish1897,hahmann1912,LingYuquan1998,LingYuquan2003,cheng_experimental_2018,Li2003,liu_experimental_2020,alvarez_ripples_2025}, and this study). In an attempt to avoid this obvious contradiction of a ripple's destruction criterion undercutting its formation criterion, one might hypothesize that with increasing free-stream wind velocity, the surface shear stress does not increase. It would rather remain pinned at the cessation threshold (the minimum shear stress that sustains aeolian saltation) due to the negative feedback of the transported sand grains on the wind \cite{kok_physics_2012}. 
As a consequence, though, this would prevent the surface shear stress from reaching the fluid-entrainment threshold, in the first place, annulling the basis on which the conventional picture explains impact ripple suppression. Ultimately, this would (wrongly) imply that impact ripples exist for any wind strength, making the inconsistency even worse. 

As a refinement of the original fluid-entrainment hypothesis, one could argue that the flux saturation lags spatially behind the shear stress variation induced by the ripple topography, leading to undersaturated flux conditions at the ripple crests. 
With increasing wind strength, the erosion would then merely lower and gradually degrade the amplitude of the impact ripples, leaving their topological pattern intact. This qualitative statement about the evolution of the ripple pattern topography during ripple disappearance can be tested in experiments and does not rely on any precise measurement of wind characteristics like $u_*$. Our experiments, however, cannot confirm it and instead support quite a different scenario. 

\subsubsection*{Main experimental finding}
To elucidate how ripple patterns actually deteriorate under strong winds, Fig.~\ref{fig:crestline_plots_w_clusters} compares impact ripple topographies produced in a wind tunnel in terms of detected crest lines, for two exemplary shear velocities (see SI~Fig.~S3 for results for all employed wind strengths). The layout of ripple crest patterns shows clear evidence of a structural deterioration and disintegration of the wave pattern upon increasing wind speed, contradicting the notion of a gradual suppression of a regular pattern. For a meaningful quantification of the structural decay of a coherent wave pattern in the sand bed, the percentage of detected peaks organized in long crestlines is employed as order parameter $P$ (see Methods). As evident from Fig.~\ref{fig:crestline_plots_w_clusters}(c), it exhibits a drastic drop with increasing wind strength, evoking the notion of a dynamical phase transition, rather than the gradual suppression implied by the conventional understanding. Our observations, together with the outlined conceptual inconsistencies and shortcomings of the fluid-entrainment hypothesis, clearly demonstrate the need for alternative explanations of impact-ripple fragility.

\subsubsection*{Mechanisms of impact-ripple fragility}

\begin{figure}
\centering
\begin{tikzpicture}
    \node[anchor=north west, inner sep=0] (image) at (0,0) {\includegraphics[width = .88\linewidth]{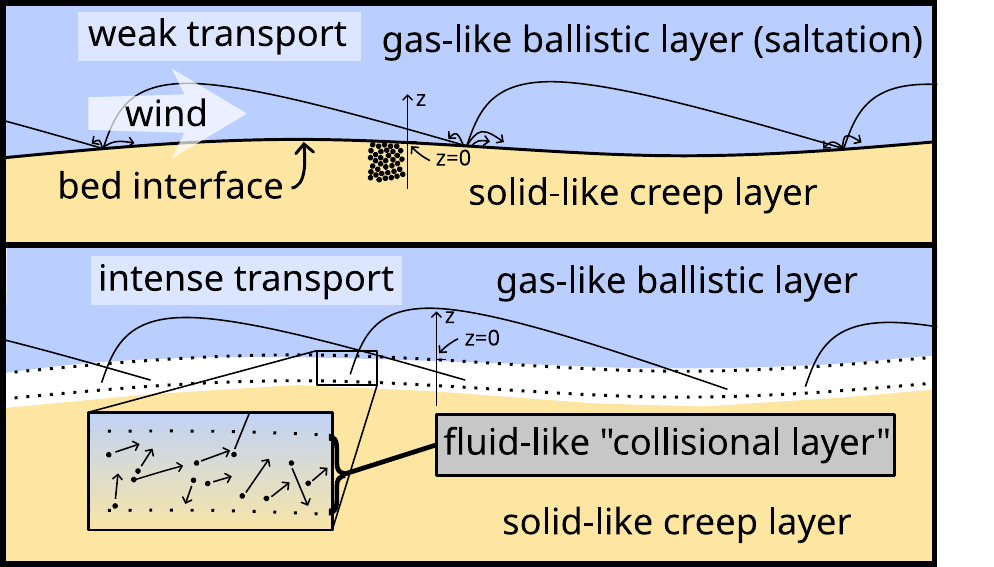}};
    \node[anchor = north east] at (-.3,0) {\textbf{a}};
\end{tikzpicture}\\
\vspace{5pt}
\begin{tikzpicture}
    \node[anchor=south west, inner sep=0] (image) at (0,0) {\includegraphics[width = \linewidth]{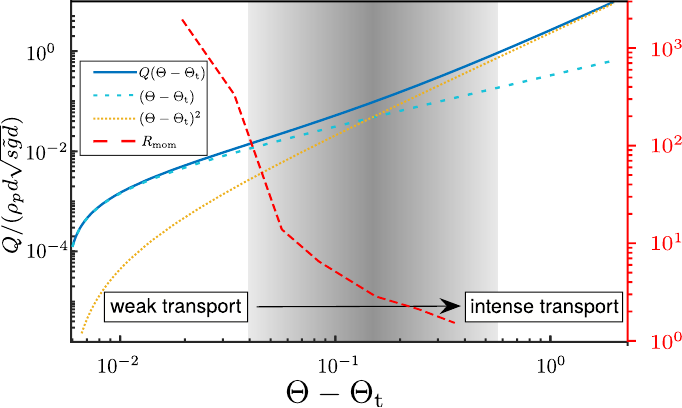}};
    \node[anchor = north west] at (image.north west) {\textbf{b}};
\end{tikzpicture}%
\caption{\textbf{Aeolian surface melting of a granular bed upon transitioning from the weak to the intense aeolian sand transport regime.} \textbf{(a)}: Schematic of aeolian sand transport regimes. The weak transport regime at low wind speed is characterized by a dilute transport layer, in which grains hop on ballistic saltation trajectories over a quasi-static, close-packed granular bed. Impacting grains  rebound from the bed surface, inducing impact-ripple formation by splash. Upon increasing the wind speed, the creeping skin layer of the bed surface ``melts'' into a fuzzy, collision-dominated third layer, blurring the interface and screening the bulk from the saltating grains. Together with the sharp decline of the scale separation between the momentum transport speeds above and below the bed interface, this impedes impact-ripple formation. \textbf{(b)}: Crossover scaling of the particle flux (left ordinate axis), quantified in terms of the sediment transport rate $Q$ as a function of the excess $\Theta - \Theta_t$ of the Shields number $\Theta = u_*^2/ (s\tilde{g}d)$ at the transition from weak to intense granular transport. Here, $u_*$  denotes the shear velocity, $d$ the grain diameter, $\rho_p$ the grain material density, $s$ the grain to air density ratio, $\tilde{g}$ the buoyancy-reduced gravitational constant (adapted from the transport law established in Ref.~\cite{Pahtz_scaling_laws_2023}). The ratio $R_\text{mom}$ (right ordinate axis) of the collisional/elastic to ballistic momentum transport speeds, estimated just below and above the nominal bed surface, respectively, drops significantly in the crossover regime and approaches unity in the intense transport regime (see Supporting Information for further details).}
\label{fig:sketch_smh}
\end{figure}

We have argued above that a consistent notion of \emph{megaripple} fragility links their decay to the breakdown of a fundamental characteristic precondition of their formation, the bimodal transport mode. Similarly, we now connect impact-ripple erasure mechanistically with a recently revealed fundamental change in the aeolian transport process for monodisperse sands \cite{carneiro_midair_2013,ralaiarisoa_transition_2020}. Namely, from ``weak'' to ``intense'' transport conditions, without and with an important role of mid-air collisions, respectively \cite{Paehtz_unification_2020,carneiro_midair_2013,ralaiarisoa_transition_2020,Tholen_scaling_2023}, as schematically depicted in Fig.~\ref{fig:sketch_smh}(a). The fundamental difference manifests itself in a crossover scaling of the (dimensionless) sediment transport rate $Q$ in the excess Shields number $\Theta - \Theta_\text{t}$, Fig.~\ref{fig:sketch_smh}(b), where $\Theta \propto u_*^2$ and $\Theta_\text{t}$ is the Shields number at the cessation threshold \cite{Paehtz_unified_2021,Pahtz_scaling_laws_2023}. Close to the threshold, where $\Theta \approx \Theta_\text{t}$, $Q$ scales linearly in $\Theta - \Theta_\text{t}$, and the height-resolved mean grain velocity exhibits a clear scale separation between particles moving above and below the bed surface, as apparent from Fig.~3 of Ref.~\cite{Tholen_scaling_2023}. This corroborates the classical picture of aeolian sand transport under terrestrial conditions as a two-layer flow: a ballistic transport layer characterized by saltation trajectories with negligible mid-air collisions flows over an inert bed, merely displacing some reptating grains uppon impacts. 

Upon entering the intense transport regime with increasing shear stress, this neat picture breaks down~\cite{Tholen_scaling_2023}. The bed interface gets increasingly blurred and is then more accurately described as a fluid-like crossover region of collision-dominated transport, sometimes denoted as ``collisional layer'', cf.~Fig.~\ref{fig:sketch_smh}(a). Which evokes the notion of a stratified three-phase flow with, figuratively speaking, a succession of a slowly creeping solid, a flowing fluid and a streaming rarefied gas. More quantitatively, increasing $u_*$ by a factor of 3 speeds up bed creep in a thin skin layer (about a grain diameter deep) by a factor of 500 \cite{Tholen_scaling_2023}. Thereby, the dynamical distinction between the propagation above and below the nominal bed surface is largely abolished. If, in analogy to continuum fluid mechanics, cast in terms of a Froude-like number $\text{Fr}$, the latter would jump from $\text{Fr} < 10^{-2}$ to $\text{Fr} \approx 1$, signaling the vanishing of surface waves. 

A complementary grain-scale perspective onto the phenomenology is provided by the characterization of the bed surface as the locus where the dominant momentum-transport mechanism changes from ballistic mass transport to collisional/elastic force propagation without noticeable mass transport. In the weak transport regime, force propagation through the preset network of close grain contacts in the bed \cite{owens_sound_2011} can by far outpace  ballistic mass transport, as familiar from a Newton cradle. But its speed depends sensitively on the granular packing and abruptly drops by orders of magnitudes, when the packing density decreases due to the dilatancy effect if the bed is sheared under intense flow conditions. Then, not only the scale separation between the ballistic grain speeds above and below the bed surface breaks down~\cite{Tholen_scaling_2023}, but also that between the corresponding momentum transport efficiencies, as indicated by the dashed red curve in Fig.~\ref{fig:sketch_smh}(b). The corresponding fluidization of the bed surface and the emergence of the third layer screen the solid bulk from the impacts of saltation trajectories, as betrayed by a decreasing rebound probability for saltating grains compared to a static bed \cite{wang_rheological_2024}. Thereby, the preconditions for any mechanism based on the deformations of a  solid sand bed by impacting grains are increasingly undermined. 

\subsubsection*{The surface melting hypothesis}
In summary, all mechanistic explanations of impact ripple formation rely on the premise that fast saltating grains directly impact on an inert ``solid'' bed \cite{Bagnold41,duran_direct_2014,lester_emergence_2025} that quickly shares the impact momentum with its bulk.  That this premise holds only under weak transport conditions  and breaks down for intense flows is  our central ``surface-melting hypothesis''.  As soon as bed dilatancy and mid-air collisions play a substantial role in the transport process, the scale separations between the slowly creeping bed and the fast granular flow above it, as well as that between the fast collisional momentum transport along a preformed contact network and the ballistic momentum transport between collisions, get increasingly blurred. The bed surface then virtually melts, and one may figuratively speak of  ``ripples drowning in quicksand''.

\subsubsection*{Further experimental validation}

\begin{figure}
\begin{center}
\includegraphics[width= 0.95\linewidth]{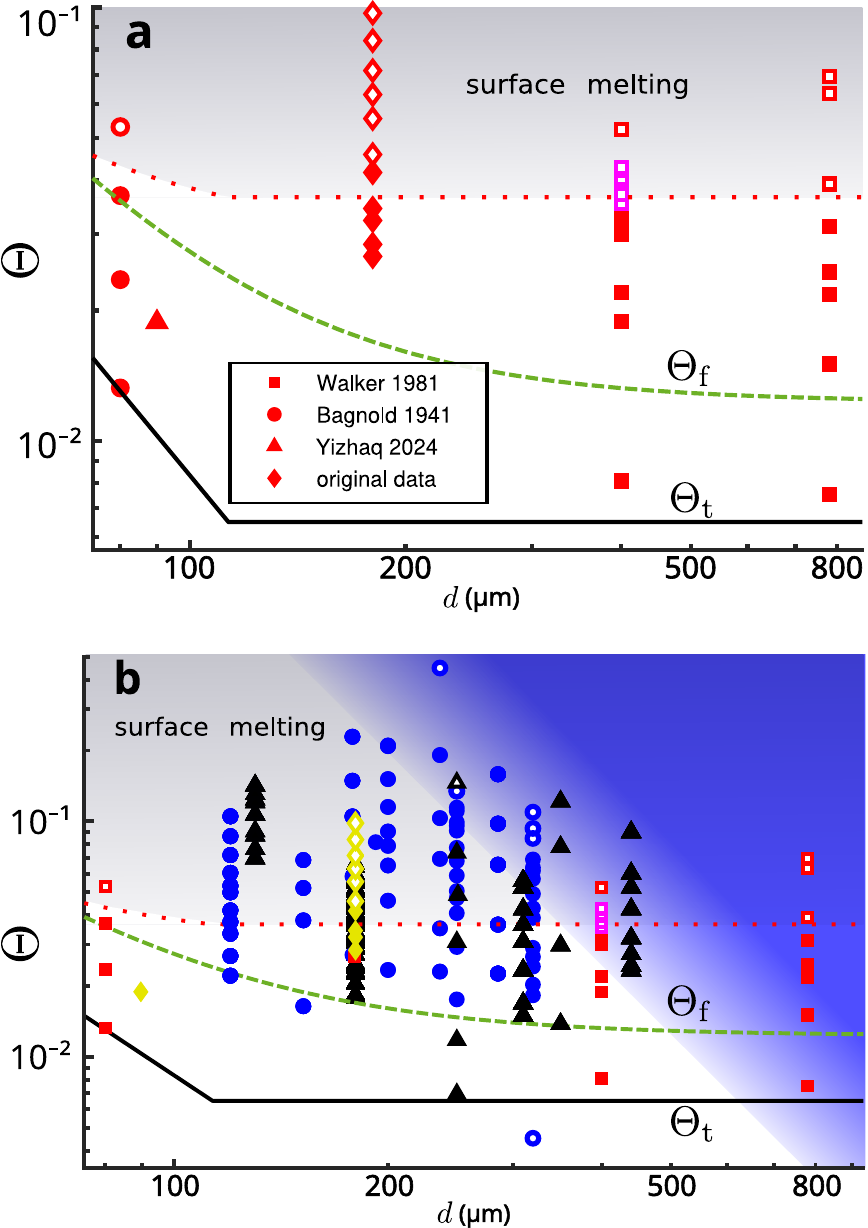}
\end{center}
\caption{\textbf{Ripple phase diagram gathering field and wind-tunnel evidence for the presence (filled symbols) and absence (empty symbols) of ripple patterns.} The dashed green and black solid lines indicate theoretical estimates of the fluid-entrainment and cessation threshold, $\Theta_\text{f}$ \cite{ShaoLu_2000} and $\Theta_\text{t}$ \cite{Pahtz_scaling_laws_2023}, respectively. The grey-shaded area indicates the crossover regime, where surface melting starts. The relative contribution of the quadratic (collisional) contribution to the sand flux $Q$ exceeds 17\% at the dotted red line. \textbf{(a)}: Impact ripple patterns disappear with the onset of surface melting. Data points above Walker's inflection point \cite{Walker81} are identified as disintegrating ripple wave trains and are marked in magenta. \textbf{(b)}: Impact ripples (red squares), aerodynamic ripples (blue circles) and non-classifiable ripples (black triangles), according to the classification in the SI. Our own original observations of co-evolving impact and aerodynamic-ripple patterns are marked with filled yellow diamonds, exclusive aerodynamic ripples (in absence of impact ripples) with open diamonds. The transition to the hydrodynamically rough regime (dark blue region) occurs at $Re_p \gtrsim \mathcal{O}(10^1)$. Ripple data originates from Refs. \cite{Walker81} (780, 400, 320, 250, 200, 440 $\unit{\micro\metre}$), \cite{Schmerler2016}(286, 237, 178 $\unit{\micro\metre}$),\cite{Bagnold41}(80, 250 $\unit{\micro\metre}$), \cite{zheng_mechanics_2009} (130 $\unit{\micro\metre}$), \cite{andreotti_aeolian_2006}(120 $\unit{\micro\metre}$), \cite{Sharp63} (310 $\unit{\micro\metre}$), \cite{Li2003} (350 $\unit{\micro\metre}$), \cite{cheng_experimental_2018} (152 $\unit{\micro\metre}$), \cite{Tholen24}(90 $\unit{\micro\metre}$), \cite{alvarez_ripples_2025} ($195 \unit{\micro\metre}$) and original experiments (diamonds at 180 $\unit{\micro\metre}$).}
\label{fig:theta_d_plots}
\end{figure}

The surface-melting hypothesis not only provides a consistent mechanistic framework to understand impact-ripple fragility. We believe that it also stands up very well to our own original data and literature data for ripples in monodisperse sand under aeolian terrestrial conditions. A direct comparison between theory and experiment is not straightforward, though. Our validation of the predicted quantitative stability diagram  in Fig.~\ref{fig:theta_d_plots} requires us to discriminate between impact ripples and aerodynamic ripples --- a subtle distinction most observers were probably not aware of when performing their measurements. 

Since the surface-melting hypothesis does not concern the ripple wavelength and height, and to keep our assumptions at a minimum, sufficiently revealing clues concerning the structural properties of the ripple patterns and/or their disappearance have to be found in the images and text of included publications. Careful assessment allows for a reasonably reliable classification of at least some of the literature data into impact and aerodynamic ripples. For a thorough discussion and justification of the classification, the reader is referred to the Supporting Information. Here, we can only summarize the main results. 

In Fig.~\ref{fig:theta_d_plots}, ripple data is displayed in $\Theta$--$d$ space, where $\Theta$ is the Shields number and $d$ is the mean grain diameter. A note on the measurements for $d=400\unit{\micro\metre}$ by Walker \cite{Walker81} is in order: at an ``inflection point'', he reports a very irregular ripple pattern, matching well the structural decay of impact ripples, as outlined above and illustrated in Fig.~2. We therefore conclude from Walker's careful description that the co-existence of impact and aerodynamic ripples was observed until impact ripples vanished at said inflection point, and that higher wind speeds produced only aerodynamic ripples. Besides the data, all relevant transport thresholds are displayed in Fig.~\ref{fig:theta_d_plots}(a). The onset of surface-melting exceeds by more than 400\% the threshold for the breakdown of bimodal transport of megaripples, located at approximately the cessation threshold $\Theta \approx\Theta_t$, which demarcates the upper bound for megaripple stability. This explains the observed survival of (supposed) impact ripples upon megaripple erasure by storms \cite{isenberg_2011,yizhaq_evolution_2012,yizhaq_morphology_2008,yizhaq_longevity_2015,john_m_ellwood_peter_d_evans_an_small_1975,Bagnold41}. Depending on the value of $d$, the critical $\Theta$ for surface melting may even exceed the fluid entrainment threshold $\Theta_\text{f}$ by around 14\% to 85\%. Which explains observations of impact ripples above the fluid-entrainment threshold \cite{Bagnold41,Walker81,Schmerler2016,Tholen24,zheng_mechanics_2009,Cornish1897,hahmann1912,LingYuquan1998,LingYuquan2003,cheng_experimental_2018,Li2003,liu_experimental_2020}. In fact, the vast majority of ripple patterns in monodisperse sands on Earth were observed above the fluid-entrainment threshold, as shown in Fig.~\ref{fig:theta_d_plots}(b), in conflict with the historical view that such winds would destroy ripples. However, consistent with the surface-melting hypothesis, our theoretically estimated onset of surface melting marks a clear cutoff in the existence of impact ripples while leaving other ripple patterns unaffected. Figure~\ref{fig:theta_d_plots} thus visualizes both a significant failure of the conventional explanation and a remarkable validation of our new hypothesis. In addition, our tentative phase diagram of Fig.~\ref{fig:theta_d_plots}(b) supports the notion that also aerodynamic ripples may vanish, albeit independently of the smaller impact ripples. The data seem consistent with their absence in the so-called aerodynamically rough regime, characterized by large particle Reynolds numbers $Re_\text{d} = u_* d/\nu \gtrsim 20$ (blue-shaded region), analogous to hydrodynamic ripples in water \cite{duran_vinent_unified_2019}. This observation underlines the hydrodynamic nature of aerodynamic ripple formation, as opposed to impact ripples, lending further independent support to the diversity of the two types of aeolian ripples and the novel ripple classification scheme of Ref.~\cite{Tholen24}.

\section*{Conclusion}
.
The surface-melting hypothesis, introduced above, not only updates our understanding of impact-ripple fragility but provides a more coherent understanding of aeolian ripples, in general. It neither suffers from conflicts with experimental data nor do its underlying physical arguments conceptually undermine the current mechanistic understanding of other ripple types.
In particular, by disentangling the respective limiting mechanisms, the observed survival of small scale (impact) ripples after megaripple flattening is naturally and unambiguously explained.

The surface-melting hypothesis is based on fundamental properties of the grain-scale aeolian sand transport process. Rather than postulating the dominance of a destructive over a competing constructive mechanism, it addresses the breakdown of the logical prerequisites (saltation impacts on an elastically intact bed) for any pertinent constructive grain-scale mechanism, themselves, without necessarily requiring its complete and detailed modelling. This underscores the robustness and predictive power of the  argument. 

Moreover, mid-air grain collisions, which have sparked many researchers' interest recently \cite{ralaiarisoa_transition_2020,Paehtz_unification_2020,Tholen_scaling_2023,carneiro_midair_2013,huo_mid-air_2024}, become thereby  associated with a pronounced qualitative and macroscopically observable effect on aeolian structure formation. As yet another added benefit, the surface-melting hypothesis provides a sound criterion for distinguishing aerodynamic ripples and impact ripples (a task that proved very difficult in the past), based on their distinct stability criteria and their dependence on environmental parameters.

Our tentative ripple phase diagram in Fig.~\ref{fig:theta_d_plots} visualizes the formation regimes of the different ripple types and validates the surface-melting hypothesis. As a challenge for future systematic investigations, it strongly suggests that the recently introduced aerodynamic ripples may only form in the aerodynamically smooth regime, characterized by moderate particle Reynolds numbers.
Field researchers looking for evidence of ripple disappearance to verify the proposed thresholds will be confronted with the relatively fast ripple formation process, resulting in a quick recovery of ripple patterns during phases of wind abatement after storms. This could explain why ripple disappearance is easily missed. And it encourages direct measurements during strong winds, which are challenging and preferably avoided during field trips. 

\section*{Methods}
\subsection*{Ambient air wind tunnel}

Experiments were conducted at the stationary-boundary-layer wind tunnel of the Aeolian Simulation Laboratory at Ben-Gurion University, Israel. Details of the experimental setup can be found in Refs. \cite{Katra2014,Pye2008}. Briefly, the test section has a cross section of $0.7 \unit{\metre} \times 0.7 \unit{\metre}$ and is $7\unit{\metre}$ long. 
The tunnel was preloaded with a $3 \unit{\centi\metre}$ deep bed of artificial spherical glass beads of mean diameter $d = 180 \unit{\micro\metre}$ and density $\rho_p = 2500 \unit{\kilogram\per\metre\tothe{3}}$, for the grain size distribution see SI~Fig.~S1. The shear velocity $u_*$ was determined from a log-fit against wind speed measurements at three different heights above the sand bed. The flat bed was exposed to stationary winds for the run times shown in SI~Table~S1. Videos were captured from the top of the wind tunnel during the experiments. The sand-bed topography was captured, after pausing the wind exposure, using the ``EinScan Pro HD'' (SHINING 3D) scanner in fixed scan mode. 

Our experiments confirm the previously observed co-evolution of impact ripples and aerodynamic ripples \cite{Tholen24} at low wind speeds (see Fig.~\ref{fig:ripple-evolution-images}(c-d)). Impact ripple profiles were recorded until shallow aerodynamic ripples started to emerge (Fig.~\ref{fig:ripple-evolution-images}(c,e)). The experiment was then continued to analyze aerodynamic ripples with more pronounced height profiles(Fig.~\ref{fig:ripple-evolution-images}(d,f)).  

The ripple wavelengths and heights were extracted from the measured height profiles of the sand surface with a horizontal resolution of $1\unit{\milli\metre} \times 1\unit{\milli\metre}$. Ripple peaks and troughs where detected with the help of smoothing splines from two-dimensional cuts along the wind direction through the three-dimensional data. The ripple characteristics deduced from our experiments are summarized in SI~Table~S1. The aerodynamic ripple features match very well the previously reported characteristics of ripples produced in the same wind tunnel from an almost identical grain size distribution \cite{Schmerler2016}. Impact ripple wavelengths were found to only slightly increase with increasing shear velocity, their heights are on the order of the grain diameter $d$.

The order parameter $P$ is defined as the percentage of crests that belong to connected crest lines longer than a suitable threshold length of $20\unit{\milli\metre}$. A connected crest line is defined as a set of neighbouring peaks with nearest-neighbour distance below $2\unit{\milli\metre}$ (around $11 d$). The length of connected crest lines is estimated by the summed nearest-neighbour distances. Besides representative data in Fig.~\ref{fig:crestline_plots_w_clusters}, SI~Fig.~S3 provides an overview of our complete experimental data.

\printbibliography

\subsubsection*{Acknowledgements}
This research was supported by the the German-Israeli Foundation for Scientific Research and Development (GIF, grant no. I-155-301.10-2018), and by the Israel Science Foundation (ISF, grant no. 1945/25).
.

\subsubsection*{Competing Interests}
The authors declare no competing interests.

\subsection*{Data Availability}

\subsection*{Additional Information}
Supplementary Information is available for this paper.\\
Correspondence and requests for materials should be addressed to Constantin Rein.
\end{document}